\documentclass[twoside,reqno]{lt7-proc}
\usepackage{epsfig,cite}
\usepackage{amssymb,amsmath}
\usepackage{times}
\begin{document}
\sloppy \raggedbottom
\setcounter{page}{1}


\newcommand{\nc}{\newcommand}

\nc{\be}{\begin{equation}}

\nc{\ee}{\end{equation}}

\nc{\bea}{\begin{eqnarray}}

\nc{\eea}{\end{eqnarray}}

\nc{\xx}{\nonumber\\}

\nc{\ct}{\cite}

\nc{\la}{\label}

\nc{\eq}[1]{(\ref{#1})}


\def\IB{{\hbox{{\rm I}\kern-.2em\hbox{\rm B}}}}
\def\IC{\,\,{\hbox{{\rm I}\kern-.50em\hbox{\bf C}}}}
\def\ID{{\hbox{{\rm I}\kern-.2em\hbox{\rm D}}}}
\def\IF{{\hbox{{\rm I}\kern-.2em\hbox{\rm F}}}}
\def\IH{{\hbox{{\rm I}\kern-.2em\hbox{\rm H}}}}
\def\IN{{\hbox{{\rm I}\kern-.2em\hbox{\rm N}}}}
\def\IP{{\hbox{{\rm I}\kern-.2em\hbox{\rm P}}}}
\def\IR{{\hbox{{\rm I}\kern-.2em\hbox{\rm R}}}}
\def\IZ{{\hbox{{\rm Z}\kern-.4em\hbox{\rm Z}}}}


\def\CA{{\cal A}}
\def\CC{{\cal C}}
\def\CD{{\cal D}}
\def\CE{{\cal E}}
\def\CF{{\cal F}}
\def\CG{{\cal G}}
\def\CH{{\cal H}}
\def\CK{{\cal K}}
\def\CL{{\cal L}}
\def\CM{{\cal M}}
\def\CN{{\cal N}}
\def\CO{{\cal O}}
\def\CP{{\cal P}}
\def\CR{{\cal R}}
\def\CS{{\cal S}}
\def\CU{{\cal U}}
\def\CV{{\cal V}}
\def\CW{{\cal W}}
\def\CY{{\cal Y}}
\def\CZ{{\cal Z}}


\def\half{\frac{1}{2}}
\def\p{\partial}


\def\vare{\varepsilon}
\def\zbar{\bar{z}}
\def\wbar{\bar{w}}
\def\what#1{\widehat{#1}}


\newpage
\setcounter{figure}{0}
\setcounter{equation}{0}
\setcounter{footnote}{0}
\setcounter{table}{0}
\setcounter{section}{0}



\title{Noncommutative Spacetime and Emergent Gravity}

\runningheads{H. S. Yang}{Noncommutative Spacetime \& Emergent Gravity}

\begin{start}


\author{Hyun Seok Yang}{1,2}

\address{School of Physics, Korea Institute for Advanced Study, Seoul 130-012,
Korea}{1} \\
\address{Institut f\"ur Physik, Humboldt Universit\"at zu Berlin,
Newtonstra\ss e 15, D-12489 Berlin, Germany}{2}


\begin{Abstract}
We argue that a field theory defined on noncommutative (NC) spacetime should be regarded as
a theory of gravity, which we refer to as the emergent gravity.
A whole point of the emergent gravity is essentially originated from the basic property:
A NC spacetime is a (NC) phase space. This fact leads to two important consequences: \\
(I) A NC field theory can basically be identified with a matrix model or a large $N$ field theory
where NC fields can be regarded as master fields of large $N$ matrices. \\
(II) NC fields essentially define vector (tetrad) fields. So they define a gravitational metric
of some manifold as an emergent geometry from NC gauge fields. \\
Of course, the pictures (I) and (II) should refer to the same physics, which should be familiar
with the large $N$ duality in string theory. The $1/N$ corrections in the picture (I) correspond to
the derivative corrections in terms of the noncommutativity $\theta$ for the picture (II).

\end{Abstract}
\end{start}


\section{Noncommutative spacetime and large N duality}

Quantum mechanics is the formulation of mechanics on noncommutative (NC) phase space
\be \la{quantum-mechanics}
[x^a,p_b] = i \hbar {\delta^a}_b.
\ee
The noncommutativity of phase space leads to the Heisenberg's uncertainty relation, e.g.,
$\Delta x \Delta p \geq \hbar / 2$ which indicates that we need a large energy to probe
short distances. The large energy localized at short distances will deform a background geometry
according to the equivalence principle. This kind of backreaction of the background spacetime
will introduce a new kind of uncertainties. This new uncertainty appears as a NC spacetime \ct{DFR}.
In other words, quantum mechanics necessarily implies NC geometry at short distances.

A NC spacetime $M$ is obtained by introducing a symplectic
structure $B = \half B_{ab} dy^a \wedge dy^b$ and then by quantizing
the spacetime with its Poisson structure $\theta^{ab} \equiv (B^{-1})^{ab}$,
treating it as a quantum phase space. That is, for $f,g \in C^\infty(M)$,
\be \la{poisson-bracket}
\{f, g\} = \theta^{ab} \left(\frac{\p f}{\p y^a} \frac{\p g}{\p y^b}
- \frac{\p f}{\p y^b} \frac{\p g}{\p y^a}\right) \Rightarrow
- i[\what{f},\what{g}].
\ee
According to the Weyl-Moyal map \ct{nc-review}, the NC algebra of operators
is equivalent to the deformed algebra of functions defined
by the Moyal $\star$-product, i.e.,
\begin{equation}\label{star-product}
\what{f} \cdot \what{g} \cong (f \star g)(y) = \left.\exp\left(\frac{i}{2}
\theta^{ab} \partial_{a}^{y}\partial_{b}^{z}\right)f(y)g(z)\right|_{y=z}.
\end{equation}
Through the quantization rules \eq{poisson-bracket} and \eq{star-product},
one can define NC $\IR^{2n}$ by the following commutation relation
\be \la{nc-spacetime}
[y^a, y^b]_\star = i \theta^{ab}.
\ee
For simplicity we will consider Euclidean NC spaces with constant $\theta^{ab}$.

\newpage

A simple but crucial observation is that the NC spacetime \eq{nc-spacetime} is actually
a (NC) phase space where $\theta^{ab}$ defines its Poisson structure. This fact leads to two
important consequences \ct{hsy5,hsy6}.

(I) If we consider a NC $\IR^2$ for simplicity,
any field $\widehat{\phi} \in \CA_\theta $ on the
NC plane can be expanded in terms of the complete operator basis
\begin{equation}\label{matrix-basis}
\CA_\theta = \{ |m \rangle \langle n|, \; n,m = 0,1, \cdots \},
\end{equation}
that is,
\begin{equation}\label{op-matrix}
    \widehat{\phi}(x,y) = \sum_{n,m} M_{mn} |m \rangle \langle n|.
\end{equation}
One can regard $M_{mn}$ in \eq{op-matrix} as components of an $N \times N$
matrix $M$ in the $N \to \infty$ limit.
We then get the following relation:
\be \la{sun-sdiff}
\mathrm{Any \; field \; on \; NC} \; \IR^2 \; \Longleftrightarrow  N \times N \;
\mathrm{matrix} \; \mathrm{at} \; N \to \infty.
\ee
If $\widehat{\phi}$ is a real field, then $M$ should be a Hermitian matrix.
The relation \eq{sun-sdiff} means that NC fields can be regarded
as master fields of large $N$ matrices \ct{master}.

(II) An important fact is that translations in NC directions are an inner automorphism of
NC C*-algebra $\CA_\theta$, i.e.,
$e^{ik \cdot y} \star f(y) \star e^{-ik \cdot y} = f(y +  \theta \cdot k)$
for any $f(y) \in \CA_\theta$ or, in its infinitesimal form,
\be \la{inner-der}
[y^a, f]_\star = i \theta^{ab} \p_b f.
\ee
In the presence of gauge fields, the coordinates $y^a$
should be promoted to the covariant coordinates defined by
\begin{equation}\label{cov-coord}
 x^a (y) \equiv y^a + \theta^{ab} \widehat{A}_b (y)
\end{equation}
in order for star multiplications to preserve the gauge covariance \ct{madore}.
The inner derivations \eq{inner-der} are accordingly covariantized
too as follows
\bea \la{vector-map}
{\rm ad}_{x^a}[f] &\equiv& [x^a, f(y)]_\star = i\theta^{\alpha\beta}
\frac{\partial x^a}{\partial y^\alpha}\frac{\partial f}{\partial y^\beta} + \cdots \xx
&\equiv& V_a^\alpha(y) \partial_\alpha f(y) + {\cal O}(\theta^3).
\eea
It turns out that the vector fields $V_a (y) \equiv V_a^\alpha (y) \partial_\alpha$
form an orthonormal frame and hence define vielbeins of a
gravitational metric \ct{hsy3}.

Two remarks are in order. We may notice that the pictures (I) and (II) should refer
to the same physics, which is essentially an equivalent statement with the large $N$
duality in string theory.
The other remark is that $1/N$ corrections in the picture (I) correspond to
NC deformations in terms of $\theta$ for the picture (II)
since $A \sim \theta N$, for example, in two dimensions because $\theta$ is
a unit area and $N$ is a number of states on the plane of area $A$.

\section{DBI action as a generalized geometry}

In order to understand the origin of the emergent gravity,
one has to identify the origin of diffeomorphism symmetry,
which is the underlying local symmetry of gravity.
It turned out \ct{hsy4} that the emergent gravity is deeply
related to symplectic geometry. In particular, the Darboux theorem in symplectic
geometry plays the same role as the equivalence principle
in general relativity. The Darboux theorem states that for $[\omega^\prime]=[\omega] \in H^2(M)$ where
$\omega^\prime = \omega + dA, \; \exists \; \phi: M \to M \in
Diff(M)$ such that
\begin{equation} \label{darboux}
\frac{\partial x^\alpha}
   {\partial y^a} \frac{\partial x^\beta}{\partial
    y^b} \omega^\prime_{\alpha\beta}(x) = \omega_{ab}(y)
\end{equation}
where the coordinate transformation $\phi$ is given by Eq.\eq{cov-coord}.
The local equivalence \eq{darboux} between symplectic structures leads to
a remarkable identity between DBI actions \ct{cornalba}:
\begin{equation} \label{mirror}
\int d^{p+1} x \sqrt{\det(g + \kappa (B + F(x)))}
= \int d^{p+1} y \sqrt{\det(\kappa B + h(y))}.
\end{equation}
Note that fluctuations of gauge fields now appear as an induced metric
on the brane given by
\begin{equation} \label{induced-metric}
h_{ab}(y) =  \frac{\partial x^\alpha}{\partial y^a}
\frac{\partial x^\beta}{\partial y^b} g_{\alpha\beta}.
\end{equation}

Let us consider the triple $(M, g, B)$ as the data of D-branes,
that is a derived category in mathematics, where $M$ is a smooth manifold equipped with
a metric $g$ and a symplectic structure $B$. The DBI action \eq{mirror} shows that
the triple comes into the combination  $(M, g, B)=(M, g + \kappa B)$. Thus
the `D-manifold' defined by the triple $(M, g, B)$
describes a generalized geometry \ct{generalized-geometry}
which continuously interpolates
between a symplectic geometry $(|\kappa Bg^{-1}| \gg 1)$ and
a Riemannian geometry $(|\kappa Bg^{-1}| \ll 1)$. The decoupling limit
considered in \ct{sw} corresponds to the former.

More closely, if $M$ is a complex manifold whose complex structure is given by $J$,
we see that dynamical fields in the LHS of Eq.\eq{mirror} act only as
the deformation of symplectic structure $\Omega (x) = B + F(x)$ in the triple $(M, J, \Omega)$,
while those in the RHS of Eq.\eq{mirror} appear only as the deformation of complex structure
$J^\prime(y)$ in the triple $(M^\prime, J^\prime, B)$ through the metric \eq{induced-metric}.
In this notation, the identity \eq{mirror} can thus be written as follows
\be \la{h-mirror}
(M, J, \Omega) \cong  (M^\prime, J^\prime, B).
\ee
The equivalence \eq{h-mirror} is very reminiscent of the homological mirror symmetry \ct{kontsevich},
stating the equivalence between the category of A-branes
(derived Fukaya category corresponding to the triple $(M, J, \Omega)$)
and the category of B-branes (derived category of coherent sheaves corresponding to the triple
$(M^\prime, J^\prime, B)$).

\section{Emergent gravity from noncommutative field theory}

The correspondence between NC field theory and gravity outlined in Sections 1 and 2 can be
concretely confirmed, at least, for the self-dual sectors of NC gauge theories.
Recently we showed in \ct{hsy3} that self-dual electromagnetism in
NC spacetime is equivalent to self-dual Einstein gravity.
For example, $U(1)$ instantons in NC spacetime are actually gravitational
instantons \ct{hsy12}.

The emergent gravity from NC gauge theories can be more clarified by systematically applying
to the NC gauge theories the pictures (I) and (II) in Section 1. Let us briefly summarize
the construction in \ct{hsy6}. We refer to \ct{hsy6} for more details.
Here we will assume the Minkowski signature for commutative space parts.

Consider a NC $U(1)$ gauge theory on $\IR^D = \IR^d_C \times
\IR^{2n}_{NC}$, where $D$-dimensional coordinates $X^M \; (M=1,\cdots, D)$ are
decomposed into $d$-dimensional commutative ones, denoted as $z^\mu \;
(\mu=1, \cdots, d)$ and $2n$-dimensional NC ones, denoted as $y^a \;
(a = 1, \cdots, 2n)$, satisfying the relation \eq{nc-spacetime}.
Likewise we decompose $D$-dimensional gauge fields as follows:
$A_M (z,y) =(A_\mu, \Phi^a) (z,y)$ where $\Phi^a (z,y) \equiv x^a(z,y)/\kappa$ are adjoint Higgs fields
defined by the covariant coordinates \eq{cov-coord}. One can show that,
adopting the matrix representation \eq{op-matrix}, the NC $U(1)$ gauge theory
on $\IR^d_C \times \IR^{2n}_{NC}$ is exactly mapped to the $U(N  \to \infty)$ Yang-Mills theory
on $d$-dimensional commutative space $\IR^d_C$. For example, the 10-dimensional NC $U(1)$ gauge theory
on $\IR^{1,3}_C \times \IR^{6}_{NC}$ is equivalent to the bosonic part of 4-dimensional $\CN =4$
$U(N)$ Yang-Mills theory.

According to the map \eq{vector-map}, the $D$-dimensional NC $U(1)$ gauge fields
$A_M (z,y) =(A_\mu, \Phi^a) (z,y)$ can be regarded as gauge fields on $\IR^d_C$
taking values in the Lie algebra of volume-preserving vector fields on
a $2n$-dimensional manifold $X$, i.e., the gauge group $G = SDiff(X)$:
\be \la{local-vector}
A_\mu(z) = A_\mu^a(z,y) \frac{\p}{\p y^a}, \qquad
\Phi_a(z) = \Phi_a^b(z,y) \frac{\p}{\p y^b}.
\ee
It turns out \ct{ward} that $f^{-1}(D_1, \cdots, D_d, \Phi_1, \cdots, \Phi_{2n})$ with
$D_\mu = \p_\mu - i A_\mu(z)$ forms an orthonormal frame and hence defines a metric
on ${\bf R}^d_C \times X$ with a volume form $\nu = d^d z \wedge \omega$
where $f$ is a scalar, a conformal factor, defined by
$f^2 = \omega (\Phi_1, \cdots, \Phi_{2n})$:
\begin{equation} \label{Ward}
ds^2 = f^2 \eta_{\mu\nu} dz^\mu dz^\nu + f^2 \delta_{ab}
V^a_c V^b_d (dy^c - {\bf A}^c) (dy^d - {\bf A}^d)
\end{equation}
where ${\bf A}^a = A^a_\mu dz^\mu$ and $V^a_c \Phi_b^c = \delta^a_b$.

\newpage

Note that the emergent gravity specifies only a conformal class of metrics whose specific form
depends on the choice of the volume form $\omega$, determined by a particular background.
For example, for the above 10-dimensional case, the vacuum geometry emerging from
$(A_\mu, \Phi^a) = (0, y^a/\kappa)$, is a flat $\IR^{1,9}$ for
$\omega = dy^1 \wedge \cdots \wedge dy^6$ while it is $AdS_5 \times {\bf S}^5$
for $\omega = dy^1 \wedge \cdots \wedge dy^6/\rho^2$ where $\rho^2 = \sum_{a=1}^6 y^a y^a$.
When turning on the fluctuations $(A_\mu, \Phi^a)$,
the vacuum geometry, $\IR^{1,9}$ or $AdS_5 \times {\bf S}^5$, will be deformed
and its resulting metric will be given by Eq.\eq{Ward}. In this sense, the Ward metric \eq{Ward}
describes a kind of bubbling geometry \ct{hsy6}.

\section{Conclusion and outlook}

Emergent gravity from NC field theories reveals several remarkable pictures.
It portends that gravity may be not a fundamental force but a collective
phenomenon emerging from NC (or non-Abelian) gauge fields. (Although we are here confined
to NC $U(1)$ gauge theories, it was recently suggested \ct{harold} that
a NC $U(n)$ gauge theory should be interpreted as an $SU(n)$ gauge theory coupled to gravity.)

If so, the followings are just corollaries:
Spacetime is also emergent from gauge field interactions \ct{seiberg}.
Especially a flat spacetime emerges from vacuum energy, previously identified
with the cosmological constant. This fact may resolve the long standing
cosmological constant problem \ct{hsy7}.
As a consequence, Lorentz symmetry is also emergent.

A natural question is then why spacetime at large scales is four dimensions.
If gravity emerges from gauge theories, we may notice that electromagnetism is only
a long range force in Nature so it should determine the large scale structure
of spacetime. Then, for entertainment, let us compare the number of physical polarizations of
photons and gravitons in $D$-dimensions: $A_\mu = D-2 = \frac{D(D-3)}{2}
= g_{\mu\nu}$ $\Rightarrow \; D= 1$ or $D= 4$.
Is there a deep meaning or just an incident ?



%


\section*{Acknowledgments}

We would like to thank Harald Dorn, Mario Salizzoni and
Alessandro Torrielli for invaluable discussions and also the QFT group members, especially,
Jan Plefka, of the Institut f\"ur Physik, Humboldt Universit\"at zu Berlin
for their cordial hospitality.
This work was supported by the Alexander von Humboldt Foundation.


\end{document}